\documentclass[preprint,showpacs,preprintnumbers,amsmath,amssymb,aps,pre]{revtex4-1}
\usepackage{graphicx}
\usepackage{bm}
\usepackage{dcolumn}
\usepackage{color}
\usepackage[latin1]{inputenc}
\usepackage{hyperref}
\hypersetup{linktoc=page}
\hypersetup{
    colorlinks,
    citecolor=blue,    filecolor=blue,
    linkcolor=blue,     urlcolor=blue
}

\begin{document}

\title{Pair creation in collision of $\gamma$--ray beams produced with high intensity lasers}

\author{X. Ribeyre}
\email{ribeyre@celia.u-bordeaux1.fr}
\affiliation{Univ. Bordeaux-CNRS-CEA, Centre Lasers Intenses et Applications, UMR 5107, 33405 Talence, France} 
\author{E. D'Humi\`eres}
\affiliation{Univ. Bordeaux-CNRS-CEA, Centre Lasers Intenses et Applications, UMR 5107, 33405 Talence, France} 
\author{M. Lobet}
\affiliation{Univ. Bordeaux-CNRS-CEA, Centre Lasers Intenses et Applications, UMR 5107, 33405 Talence, France} 
\affiliation{CEA, DAM, DIF, F-91297, Arpajon, France} 
\author{O. Jansen}
\affiliation{Univ. Bordeaux-CNRS-CEA, Centre Lasers Intenses et Applications, UMR 5107, 33405 Talence, France} 
\author{S. Jequier}
\affiliation{Univ. Bordeaux-CNRS-CEA, Centre Lasers Intenses et Applications, UMR 5107, 33405 Talence, France} 
\author{V. T. Tikhonchuk}
\affiliation{Univ. Bordeaux-CNRS-CEA, Centre Lasers Intenses et Applications, UMR 5107, 33405 Talence, France} 
\date{\today}

\begin{abstract}
Direct production of electron--positron pairs in two photon collisions, the Breit--Wheeler process, is one of the basic processes in the Universe. However, it has never been observed in laboratory because of absence of the intense gamma-ray sources. Laser induced synchrotron sources emission may open for the first time a way to observe this process. A feasibility of an experimental set--up using a MeV photon source is studied in this paper. We compare several $\gamma$--ray sources, estimate the expected number of electron--positron pairs and competing processes by using numerical simulations including quantum electrodynamic effects.  
\end{abstract}

\maketitle

\section{Introduction}\label{sec1}
According to the theory of quantum electrodynamics (QED) \cite{QED_Book}, an electromagnetic radiation with a sufficiently high energy density may create a matter in the form of particle--anti--particle pairs. The electron--positron production  $\gamma+\gamma \to e^++ e^-$ is the lowest threshold process in the photon--photon interaction, which is of crucial importance in nature, controlling the energy release in gamma ray bursts, active galactic nuclei, black holes and other explosive phenomena  \cite{Piran_2004, Ruffini_2010}. It is also responsible for the TeV cutoff in the photon energy spectrum of extra-galactic sources \cite{Nikishov_1962}. 

The $e^+ e^-$ pair creation in a collision of two photons was first theoretically predicted by Breit and Wheeler \cite{Breit_1934} following the discovery of the positron by Anderson \cite{Anderson_1933}. The effective cross section of such a process  is of the same order as the Thomson cross section, i.e. $\sim r_e^2 \sim 10^{-25}$ cm$^2$ where $r_e=e^2/4\pi \epsilon_0 m_e c^2=2.8\times 10^{-13}$\,cm the electron classical radius, $\epsilon_0$ is the vacuum dielectric permittivity and $e$ is the elementary charge. While the pair creation in photon collisions takes place on astrophysical scales \cite{Ruffini_2010}, its experimental observation is difficult because of available photon fluxes of very low intensity \cite{Burke_1997,Pike_2014}. The linear Breit--Wheeler (BW) process, $\gamma' + \gamma \to e^+ + e^-$ is the first order perturbative QED process, which is followed by the multiphoton processes $\gamma' + n\gamma \to e^+ + e^-$ \cite{Nikishov_1964, Narozhny_1965}. This multiphoton electron-positron 
pair production has been observed experimentally at the Stanford Linear Accelerator Center (SLAC) \cite{Burke_1997, Bamber_1999} in collisions of a high energy electron beam with a terawatt laser pulse. However, the electron beam energy was not sufficient for the first order BW process.
 
The SLAC experiment consisted in injecting a beam of $\sim 10^9$ electrons with an energy of 46.6\,GeV into an intense laser beam with a relativistic amplitude $a_0=eE_0/m_e \omega_0 c \sim 0.5$, where  $m_e$ is the electron mass and $c$ is the light velocity, $E_0$ and $\omega_0$ are the laser electric field amplitude and frequency. It was a two step process. At the first step, the laser photons with an energy $\hbar\omega_0 \simeq 1.9$\,eV, where $\hbar$ is the Planck constant, were converted in $\sim 30$\,GeV $\gamma$-rays in the linear and nonlinear Compton backscattering processes $n\omega +e^- \to e^- + \gamma$ with $n=1$-$4$ \cite{Bula_1996}. Approximately $10^6$ high energy photons per shot have been produced. At the second step, these high--energy photons were collided with the laser photons producing pairs $\gamma + n\omega \to e^+ + e^-$. Although the probability of this process was very low, the authors observed about 100 positrons in 20000 laser shots. According to the energy conservation,  at 
least four optical photons, $n=4$, were needed for this non--linear BW process. In this configuration, colliding with an optical laser beam the first order BW process would require 200\,GeV photons. 

Observation of the BW process is difficult because of other pair--production reactions in charged particle collisions. The major competing processes are: the electrons collision with a nucleus, $e^- + Z \to Z+ e^+ + 2e^-$, the so called ``Trident'' process, with the effective cross section $\sim Z^2\alpha^2 r_e^2$ and the Bethe--Heitler process \cite{Bethe_1934}  $\gamma + Z \to Z+ e^+ + e^-$, which has a larger cross section  $\sim Z^2\alpha r_e^2$. Here, $\alpha= e^2/4\pi \epsilon_0 \hbar c=1/137$ is the fine structure constant. Both processes are efficient for the positron production with intense laser pulses and high--Z targets \cite{Liang_1998, Myatt_2009}. However, they introduce strong limitations on the noise level for detecting the BW process. Experiments carried out on the Jupiter and OMEGA EP laser facilities \cite{Chen_2009, Chen_2014} showed production of $\sim10^{10}$ positrons per shot from a thick gold target irradiated with a laser pulse having intensity $\sim10^{20}$ W/cm$^2$. Recently, on 
the ASTRA--GEMINI laser facility, a high--density ($10^{16}$ cm$^{-3}$) and small divergence (10--20 mrad) positron beam has been created, by a high intensity laser beam irradiating a gas target and using a secondary high--Z target \cite{Sarri_PPCF_2013,Sarri_PRL_2013,Sarri_Nat_2015}. These examples illustrate the difficulty in detecting the BW process, which requires a clean interaction environment excluding heavy materials and prefers a collision of intense and energetic photon beams in vacuum. 

A possible experimental scheme for studies of the BW process was suggested recently by Pike \emph{et al.} \cite{Pike_2014}. The authors proposed to collide a GeV photon beam with a bath of thermal photons at a temperature of $\sim 300$\,eV. The GeV photons are supposed to be created in the Bremsstrahlung process of laser accelerated electrons in a mm--thick gold target. Laser intensities above $10^{21}$\,W/cm$^2$ are required to accelerate electrons to GeV energies, which then generate the  photons. The thermal X--rays can be produced inside a high--Z hohlraum with a separate laser pulse having energy of a few hundred kJ. Such an experimental configuration could be realized on the LMJ or NIF laser facilities \cite{NIF, LMJ} coupled to the petawatt systems PETAL and ARC, respectively. The authors expect a production of $\sim10^5$ BW pairs in a single laser shot. However, the proposed scheme cannot be operated with multiple laser shots, and it is prone to a high noise level due to the presence of a significant 
mass of heavy material. Moreover, the authors of a recent study of the Bremsstrahlung process on a petawatt laser facility with the intensity close to $10^{21}$\,W/cm$^2$ \cite{Henderson_2014} reported a photon spectrum with a relatively low effective temperature of less than 10\,MeV and a cut--off energy below 100\,MeV. The softening of the photon spectrum is explained by an efficient creation of secondary electrons in the gold target. These observations demonstrate the difficulty to create an efficient GeV photon source based on the Bremsstrahlung process.

In this paper, we propose another experimental approach for the observation of the BW process. The scheme relies on the collision of two relatively low energy (few MeV), intense photon beams. Such beams can be created by interacting intense laser pulses with thin aluminium targets or short and dense gas jets. By colliding two of them in vacuum, one would be able to produce a significant number of electron--positron pairs in a controllable way. It offers the possibility to conduct a multi--shot experiment with a reliable statistics on laser systems with pulse energies of the level of a few joules and in a low noise environment without heavy elements. We provide details of the experimental setup, analytical estimates and numerical simulations of the expected yield of reactions and possible ways to create a photon source with requested parameters.

\section{Breit--Weeler pairs production in laboratory}\label{sec2}
The energy threshold of the BW process is defined by the conservation of energy and momentum. Assuming that both, electron and positron are produced at rest in the center of mass reference system, the threshold condition writes
\begin{equation}\label{eq1}
E_{\gamma_1} E_{\gamma_2}=2m_e^2c^4/(1-\cos\phi)
\end{equation}
where $\phi$ is the angle between the colliding photons with energies $E_{\gamma_{1,2}}$ respectively. For the optimal geometry of a head--on collision, $\phi=\pi$, the product of energies of the colliding photons should be larger than 0.25\,MeV$^2$. The appropriate choice of photons depends on the available sources. In the SLAC experiment with a laser delivering $\sim2$\,eV photons in a very short pulse of $20-30$\,fs, one would need a counterpart source of a few hundred GeV photons. The only known source of  such energetic photons would be the Compton backscattering, which requires a few hundred GeV electron beam. It is produced in km--scale linear accelerators, which are major facilities requiring rather expensive preparation campaigns. Moreover, with expected number of Compton photons $\sim 10^5$, the probability of the BW process remains very small, which does not allow the direct BW process observation. 
\begin{figure}[!h]
\includegraphics[width=10cm]{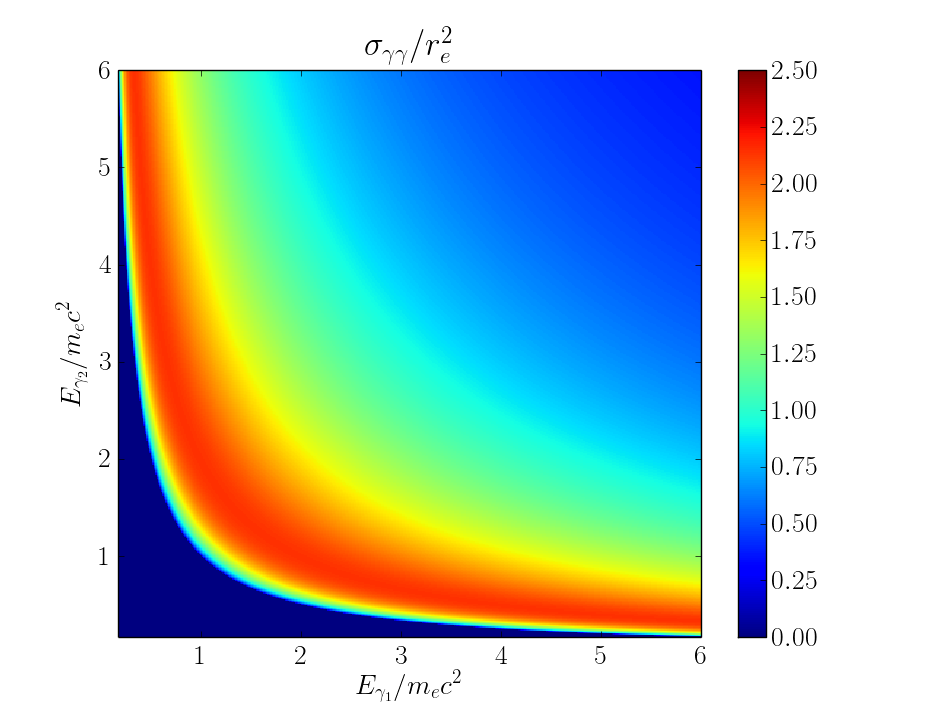} 
\caption{\label{fig1} Cross section $\sigma_{\gamma \gamma}$ versus $E_{\gamma_1}$ and $E_{\gamma_1}$ near the threshold for $\phi=\pi$.}
\end{figure}

Another known source of intense photons is a hohlraum heated by a high energy laser pulse \cite{Pike_2014}. While delivering a laser energy of a few hundred kJ inside a mm--size gold cavity, one can create a black body--like radiation with the effective temperature of $200-300$\,eV \cite{hohlraum}. This corresponds to a very significant number of photons $\sim10^{20}$ confined in a millimeter--size volume on a nanosecond time scale. The counterpart photon source is then in the GeV range, which can be created today in the laboratory--scale laser installations \cite{Leemans}. However, this scheme \cite{Pike_2014}, apart of making the photon interaction in a harsh hohlraum environment, faces another challenge in producing an efficient GeV photon source. Although several approaches could be considered \cite{Nakajima_2014}, none of them is demonstrated today, and it would be difficult to make a valid prediction of how efficient such a source could be. 

It is much easier to produce lower energy photons in a few MeV range. According to Eq. \eqref{eq1}, a collision of two such photon beams at a large angle could produce the electron--positron pairs. This is the basis of our scheme to demonstrate the BW process: a collision of two identical MeV--photon beams. Pairs production is analytically estimated in two different cases: (i) GeV photon with thermal photons bath \cite{Pike_2014} and (ii) MeV--MeV photon collision. 

The cross section of the BW process is given by the expression \cite{QED_Book}:
\begin{equation}\label{eq2}
\sigma_{\gamma \gamma}(s) =  \frac{\pi}2 r_e^2  (1-\beta^2) \left[-2\beta(2-\beta^2)+(3-\beta^4)\ln\frac{1+\beta}{1-\beta}\right],
\end{equation}
where $\beta=\sqrt{1-1/s}$ and $s=E_{\gamma_1} E_{\gamma_2}(1-\cos{\phi})/2m_e^2c^4$ is the relativistic invariant. This cross section is shown in figure \ref{fig1}. It achieves its maximum at $s\simeq 2$ and then decreases asymptotically as $1/s$. 
 
In the case (i), the thermal photons are described by the Planck distribution with the temperature $T$: $n(E_{\gamma_2})= (E_{\gamma_2}^2/\pi^2c^3\hbar^3)\,({\rm e}^{E_{\gamma_2}/T}-1)^{-1}$. By integrating the cross section \eqref{eq2} over the energy of the second photon and the collision angle, one finds a probability per unit length for a photon of the energy $E_{\gamma}$ to be converted into a pair in a collision with a thermal bath:
\begin{eqnarray}\label{eq3}
\tau_{\gamma \gamma}&=&\frac1{4\pi}\int^\infty_0 dE_{\gamma_2}\, n(E_{\gamma_2})\int \sigma_{\gamma\gamma}(s) (1-\cos\phi)\,d\Omega   \nonumber \\
&=& \frac{\alpha^2}{\pi \lambda_C}\left(\frac{T}{m_e c^2}\right)^3 F(\nu), 
\end{eqnarray}
where $\lambda_C= \hbar/ \rm m_e c$ is the Compton length and 
$$F(\nu)= \frac2{\pi r_e^2\nu^2}\int_{1/\nu}^\infty dx \,({\rm e}^x-1)^{-1}\int_0^{x\nu} s\,\sigma_{\gamma\gamma}(s)\,ds$$ 
is a function of the variable $\nu=E_{\gamma}T/m_e^2c^4$, which achieves a maximum close to unity for $\nu \simeq 2$. 

\begin{figure}
\includegraphics[width=10cm]{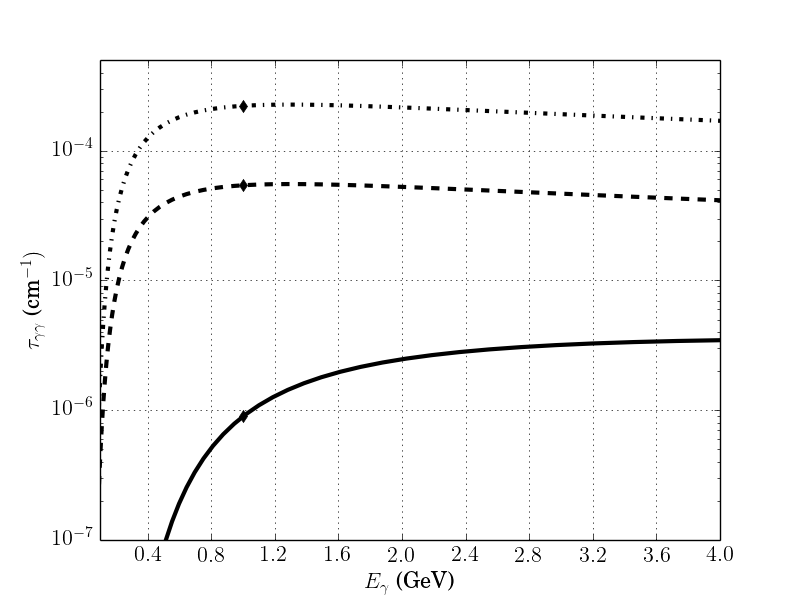}
\caption{\label{fig2} Probability of BW pair creation per unit of length versus the incident photon energy $E_\gamma$ for the thermal bath temperature, 100 (solid), 250 (dashed) and 400 eV (dash--dotted).}
\end{figure}

As one can see in figure \ref{fig2}, a probability of creating a pair with 1 GeV photon increases from $10^{-6}$\,cm$^{-1}$ to  $2\times10^{-4}$\,cm$^{-1}$, when the hohlraum temperature increases from $T=100$  to 400 eV. The total number of generated pairs, $N_p=N_{\gamma} \tau_{\gamma\gamma} L$ is proportional to the number of photons $N_\gamma$ and the propagation length $L$. Then for $L=1$\,cm and $N_{\gamma}=10^{9}$, one expects between $10^3$ and $5\times10^5$ pairs per shot in that temperature range. The lowest pair number could be small compared to the expected noise level. Moreover, in the analysis of pairs production in a 1 GeV photon interaction with thermal photon in a high--Z hohlraum one has to take into account that all pairs are generated inside the hohlraum and their detection could be difficult. The authors \cite{Pike_2014} estimate $T = 250$\,eV ($N_p=5 \times 10^4$) as a figure of merit for such an experiment. Another important parameter is the intensity of the GeV photon source: one 
needs a reliable source of $10^{9}$ photons per shot, or approximately 0.1\,J in a bunch, which is quite challenging.

In the case (ii), the use of MeV photons allows to reduce significantly the requirements on the photon source for a BW experiment. Assuming two conical $\gamma$ beams with a divergence angle $\theta$ (half angle of the full divergence) intersecting at an angle $\phi$, the interaction volume will be $V\sim 2 \pi R^2 l_\gamma (1-\cos \theta)$, where $l_\gamma= c \tau$ is the pulse length, $\tau$ is the pulse duration and $R$ is the distance between the target and the collision zone. We suppose that $R$ is much greater than the focal spot radius. The number of pairs can be estimated as $N_p\sim N_\gamma^2 \sigma_{\gamma \gamma}(\phi)/[2 \pi R^2 (1-\cos \theta )]$, where $N_{\gamma}$ is the total number of photons in the bunch. Taking for the estimate the maximum value for the cross section (\ref{eq2}) the number of pairs, for 1\,MeV beams and for $\phi=180\textdegree$ reads
\begin{equation}\label{eq4}
N_p\sim 10^8\,W^2/(R^2(1-\cos \theta)),
\end{equation} 
where $W$ is the photon beam energy in joules and $R$ is the interaction distance in $\mu$m. Therefore, two beams having an energy of 1--10\,J each, with a beam divergence angle $\theta=30\textdegree$ and an interaction distance of $R=500\,\mu$m will produce in average $3\times10^3-3\times 10^5$ pairs per shot.

\begin{figure}
\includegraphics[width=10cm]{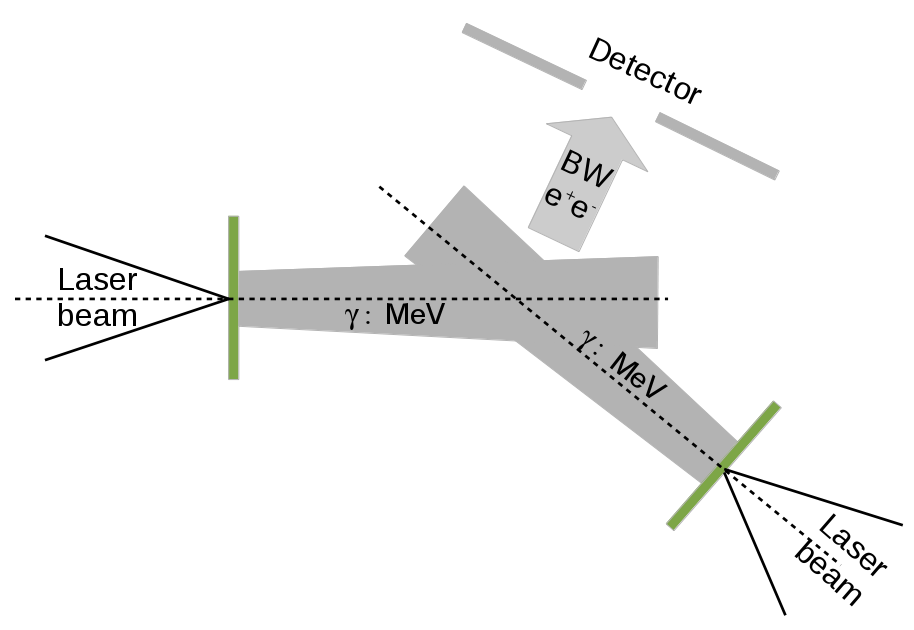}
\caption{\label{fig3} Experimental setup for the Breit--Wheeler pairs production with MeV colliding photon beams.}
\end{figure}

A source producing in average 2\,J photon bunches with an effective temperature of 6\,MeV is already available \cite{Henderson_2014}. Moreover, MeV photon bunches could be created routinely in the new generation of 10\,PW laser facilities under construction in the framework of the ELI \cite{ELI} and Apollon \cite{Apollon} projects. The schematic experimental setup is shown in figure \ref{fig3}. Two photon beams are created from thin foils irradiated with laser pulses at an high intensity of $10^{22-23}$ W/cm$^2$. A separation of the interaction zone by a distance of $1-2$\,mm should be sufficient to distinguish between the pairs created in the BW process and the background as it is shown in the next section. 

\section{Intense compact sources of MeV photons}\label{sec3}
According to the estimate \eqref{eq4} one may produce $10^3-10^4$ pairs per shot, while using beams of 1--10 J of MeV photons. Various $\gamma$--ray sources can be considered: MeV photons can be produced via Bremsstrahlung \cite{Henderson_2014}, betatron \cite{Cipiccia_2011}, Compton scattering \cite{Sarri_2014} and synchrotron emission \cite{Capdessus1}. 

The regime of emission, collimation and conversion efficiency depend on the laser intensity, duration, and the target properties. In order to make generation and detection of pairs experimentally feasible, the source should produce a collimated emission. This will allow to define the direction of the pair emission and a create sufficient intensity of the photon beam in the collision area (see Fig. \ref{fig3}). The source brightness is also a crucial parameter allowing to produce a sufficient number of pairs far from the source. Table~\ref{Tab} provides a comparison between different $\gamma$ ray sources for BW pairs production.

The photon source based on the Bremsstrahlung process can be realized by focusing an intense laser radiation on a mm--thick target of a high--Z material, gold or tungsten, for example. A large number of hot electrons is produced in the MeV energy range \cite{Compant_2014} for laser intensities $10^{20}-10^{21}\ \mathrm{Wcm}^{-2}$. In this case, 1--2 J beam  of 3--50 MeV photons with a duration of 150 fs and an average angle $\theta$ of 15\textdegree~has been reported \cite{Henderson_2014}. The laser to $\gamma$--ray  energy conversion around $2\%$. However, the use of high--Z material (gold) for the target leads to production of a large amount of background $e^+ e^-$ pairs, greater than 10$^{10}$ \cite{Chen_2009} due to the Bethe--Heitler process. This is much larger than expected number $\sim 10^4$ of BW pairs (see Table~\ref{Tab}).

The betatron sources of coherent radiation are produced with X--ray free electron lasers. By focusing a laser beam of $10^{18}$ W/cm$^2$ of a femtosecond duration inside a wave-guide plasma capillary, $10^8$ photons in range of 20--150 keV are generated with a divergence  $<$ 1\textdegree~\cite{Cipiccia_2011}. However, the efficiency of such sources decreases with the photon energy. It should be possible to produce about $10^7$ photons in range of 1--7 MeV. Then the total beam energy is $\sim 1~\mu$J, which is too low to produce a significant number of pairs per shot (see Table \ref{Tab}). 

The Thomson and Compton sources are more suited for higher photon and electron energies, above a hundred MeV, where their efficiency is higher~\cite{Nakajima_2014, Sarri_2014, Bamber_1999}. In Ref.~\cite{Sarri_2014}, one laser beam was used to produce a relativistic electron beam from a gas target, another laser beam was collided with the electron beam to produce photons from the inverse Compton scattering. In this experimental setup, $10^7$ photons at 6 MeV can be generated with a low divergence $<$ 1\textdegree.  But the pair number per shot is estimated to be $\sim 10^{-5}$, which is comparable to the betatron source (see Table \ref{Tab}). 

\begin{table}
\caption{Comparison of the different $\gamma$ sources.}\label{Tab}
\begin{ruledtabular}
\begin{tabular}{llllllll}
  Source       & Bremss.  & Betatron    & Compton    & Synch.      &\\
  \hline
  Laser energy & 100 J & 5 J & 20 J & 100 J  &\\
  $\gamma$ energy & 3--50 MeV & 1--7 MeV & 6--18 MeV  & 1--10 MeV   &\\
  Beam  energy  & 1--2 J    & 1 $\mu$J     & 1 $\mu$J & 1--10 J         &\\
  Efficiency    & $10^{-2}$&   $10^{-6}$& $10^{-7}$& $10^{-1}$ &\\
  Divergence ($\theta$) & $\sim$15\textdegree      &    $\sim$ 1\textdegree     & $\sim$1\textdegree      &    $\sim30$\textdegree       &\\ 
  Reference     &\cite{Henderson_2014}& \cite{Cipiccia_2011}&\cite{Sarri_2014}&\cite{Capdessus1} &\\    
  $N_p$ * &    $\sim 10^4$      &  $\sim 10^{-5}$        &   $\sim 10^{-5}$  &  $\sim 10^{4}$ &
\end{tabular}
\end{ruledtabular}
* Number of pairs according to Eq.(\ref{eq4}) at a distance of 500 $\mu$m.
\end{table}

In the MeV range, the most suitable source is based on the synchrotron emission of energetic electrons in an intense laser field with an intensity $\sim 10^{22-23}$~W/cm$^2$. The numerical studies predict emission of photons with energies up to tens of MeV with the conversion efficiency of several tens percent \cite{Nakamura_2014,Ridgers_2013,Ji_2014,Capdessus1}.

A promising configuration to obtain bright sources of MeV photons consists in the use of a few $\mu$m thin foil. At the laser--solid interface, the incident and reflected waves form a standing wave, producing electrons with energies up to several hundred of MeV, $\gamma_e \sim a_0$. These electrons interact with the laser field and radiate high energy photons. The characteristic energy of emitted photons $E_\gamma\sim \hbar \omega_0 a_0^3$ corresponds to the MeV range for the dimensionless laser amplitudes $a_0\sim 100$. Simulations performed in Ref. \cite{Capdessus1,Capdessus2,Capdessus3} show that tens of percent of the laser energy can be converted into a well collimated beam of $1-10$ MeV photons for the laser intensity $\sim 10^{22}$ W/cm$^2$. By colliding photons from two such sources one may produce $10^4$ pairs per shot. This number is comparable to the Bremsstrahlung source  (see Table \ref{Tab}), but with a much lower level of background pairs.

According to Table~\ref{Tab}, the Bremsstrahlung and synchrotron sources are the most suitable for pair production in the  proposed setup. 

\section{Feasibility of a MeV photon collider}\label{sec4}

With the next generation of intense laser facilities the expected laser conversion in high--energy photons is $\sim$15\%, which corresponds to an energy of $\sim$20 J, for a laser energy of 150 J as for the Apollon facility \cite{Apollon}. It is demonstrated in numerical simulations with two--dimensional particle--in--cell code CALDER~\cite{Lobet_2013}. 

\subsection{MeV photon source from solid target}

An aluminum foil 8 $\mu$m thick is irradiated at normal incidence at an intensity $10^{23}$ W/cm$^2$. The target density is 2.7 g/cm$^3$, corresponding to the number density $n_{Al} \sim 60 n_c$. The simulation is made with periodic transverse boundary conditions damping longitudinal boundary conditions both for the fields and the particles. The grid spacing is $\Delta x=\Delta y=0.03  c/\omega_0$, the time-step is equal to $\Delta t = 0.02 \omega_0^{-1}$, with 30 macro-particles per cell. The simulation domain is 300$c/\omega_0$ ($\sim 48\ \mu\mathrm{m}$) long and $45 c/\omega_0$ wide ($\sim 7\ \mu\mathrm{m}$). The pulse has a Gaussian shape with the duration $T_l = 20\pi \omega_0^{-1} \sim 33\ \mathrm{fs}$. The focal spot radius is 2 $\mu$m.

\begin{figure}
\includegraphics[width=10cm]{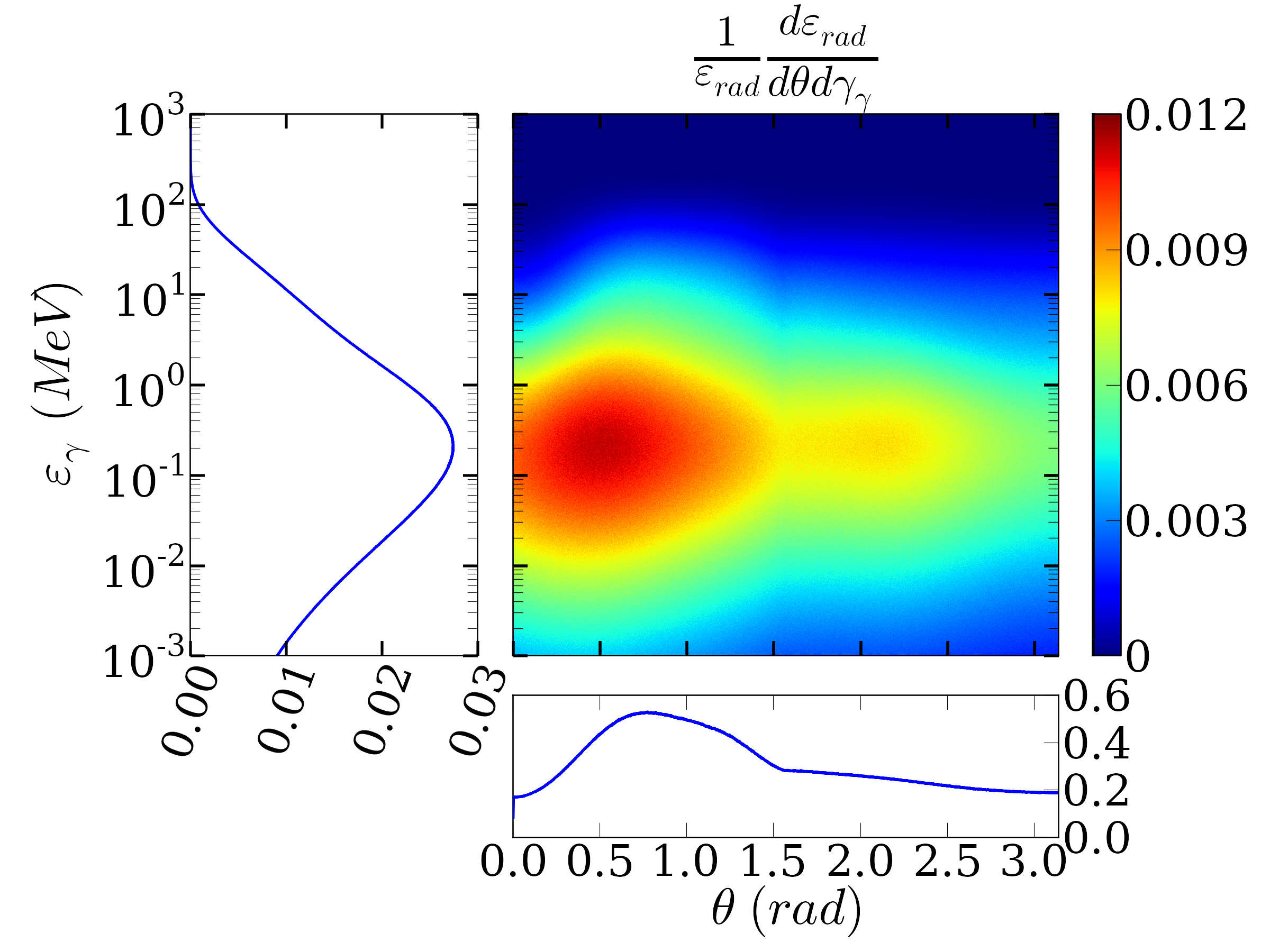}
\caption{\label{fig4} Spectral and angular distribution of the emitted photon energy in the interaction of a laser pulse at $10^{23}$ W/cm$^2$ with a 8 $\mu$m aluminium target.}
\end{figure}

A typical angular--energy spectrum of photons produced in the laser--thin foil interaction is shown in figure \ref{fig4}. The photon beam divergence is about $\sim 60$\textdegree and the maximum emission angle is at $\theta=34^\circ$ for a photon energy of 200 keV. 

The number of photons in the range of $1$--$3$ MeV emitted in a forward direction is $\sim10^{12}$. The photon brightness is $\sim$ 0.14 J/MeV/sr and the brilliance is $2\times10^{15}$ photons/sr/$\mathrm{mm}^2$/s/0.1 bandwidth.

The number of BW pairs produced in a collision of two cylindrical photon beams of densities $n_{\gamma,1}$ and $n_{\gamma,2}$ and distribution functions $f_{\gamma,1}(\gamma_1,\theta_1)$, $f_{\gamma,2}(\gamma_2,\theta_2)$ in the volume $V_\gamma = \pi R_\gamma^2 l_\gamma$ during the interaction time $T_\gamma$ reads:
\begin{eqnarray}\label{eq5}
N_{\pm,BW} & \sim & n_{\gamma,1} n_{\gamma,2} V_\gamma c T_{\gamma} \int_{\gamma_{1 min}}^{\gamma_{1 max}} \int_{0}^{\pi} f_1(\gamma_1,\theta_1) \nonumber \\ 
& & \times \int_{\gamma_{2,min}}^{\gamma_{2 max}}  \int_{0}^{\pi} { f_2(\gamma_2,\theta_2)} \sigma_{\gamma \gamma} \left( 1 - \cos{\phi} \right) d\theta_1 d\theta_2 d \gamma_{1} d\gamma_{2}
\end{eqnarray}
This integral was calculated by using the photon distribution function $f_\gamma$ shown in Fig.~\ref{fig4}. The expected number of pairs is $10^8$ produced right at the source. 

However, this configuration is not the best choice to isolate pairs produced by the Breit--Wheeler process due to the fact that the Bethe--Heitler and Trident mechanisms also contribute to the pair production. Knowing number of photons produced $\sim 10^{12}$, the expected number of pairs from the Trident process is $10^7$ for the aluminium target Z=13. 

The number of pairs produced via the Bethe--Heitler (BH) mechanism can be estimated as $N_{pBH}=\sigma_{\gamma Z} N_\gamma n_{Al} L,$ where $\sigma_{\gamma Z} \simeq Z^2 \alpha r_e^2$ the cross section $\sigma_{\gamma_Z} $, and the number of pairs produced in an aluminum target of thickness $L =8\ \mu$m is $10^9$. 

Since the expected number of BH an BW pairs is compatible at the source, the photon collision zone needs to be separated from the source. A distance of 500 $\mu$m seems to be a reasonable compromise between a reduction of the background noise and a photon beam divergence. The expected number of Breit--Wheeler pairs can be estimated from the photon function distribution shown in Fig. \ref{fig4}  according the following expression: 
\begin{eqnarray}\label{eq6}
N_{\pm,BW}  & \sim & n_{\gamma,1}n_{\gamma,2} V_{\gamma} \int_{\gamma_{2 min}}^{\gamma_{1 max}} \int_{\gamma_{2 min}}^{\gamma_{2 max}}  f_{\gamma,1} f_{\gamma,2}  \sigma_{\gamma \gamma} l_\gamma d\gamma_1 d\gamma_2.
\end{eqnarray}
At a distance of 500 $\mu$m, considering a density of $n_{\gamma} \sim 3\times 10^{19}$ cm$^{-3}$ and a divergence angle of $35^{\circ}$ we obtain a pair yield of $10^3$, which is in agreement with Eq.~\eqref{eq4} for the photon beam energy of 20~J. 

\subsection{MeV photon source from gas target}

The laser interaction with a near--critical plasma could be also a bright source of high--energy photons. Moreover by choosing a low-Z gas the yield from BH process may be suppressed. We have performed numerical simulations considering a hydrogen plasma of a density $4 n_c$ and a thickness  $l_H = 80\ \mu$m. The laser pulse at an intensity of $10^{23}\ \mathrm{Wcm}^{-2}$ ($a_0 = 270$) with a Gaussian temporal profile  $T_l$=30 fs was focused in a focal spot of a radius of 5 $\mu$m. The simulation box is 1400 $c/\omega_0$ long and 500 $c/\omega_0$ wide with a grid spacing $\Delta x=\Delta y=0.2~c/\omega_0$ and a time step $\Delta t=0.2~\omega_0^{-1}$ with 40 macro particle per cell. The plasma has a $cos^2$ density profile in the longitudinal direction and a constant in transverse direction with an initial temperature of 100 eV. Absorbing boundary conditions are applied in the longitudinal direction and thermalizing boundary conditions are used in the transverse direction.

\begin{figure}
\includegraphics[width=10cm]{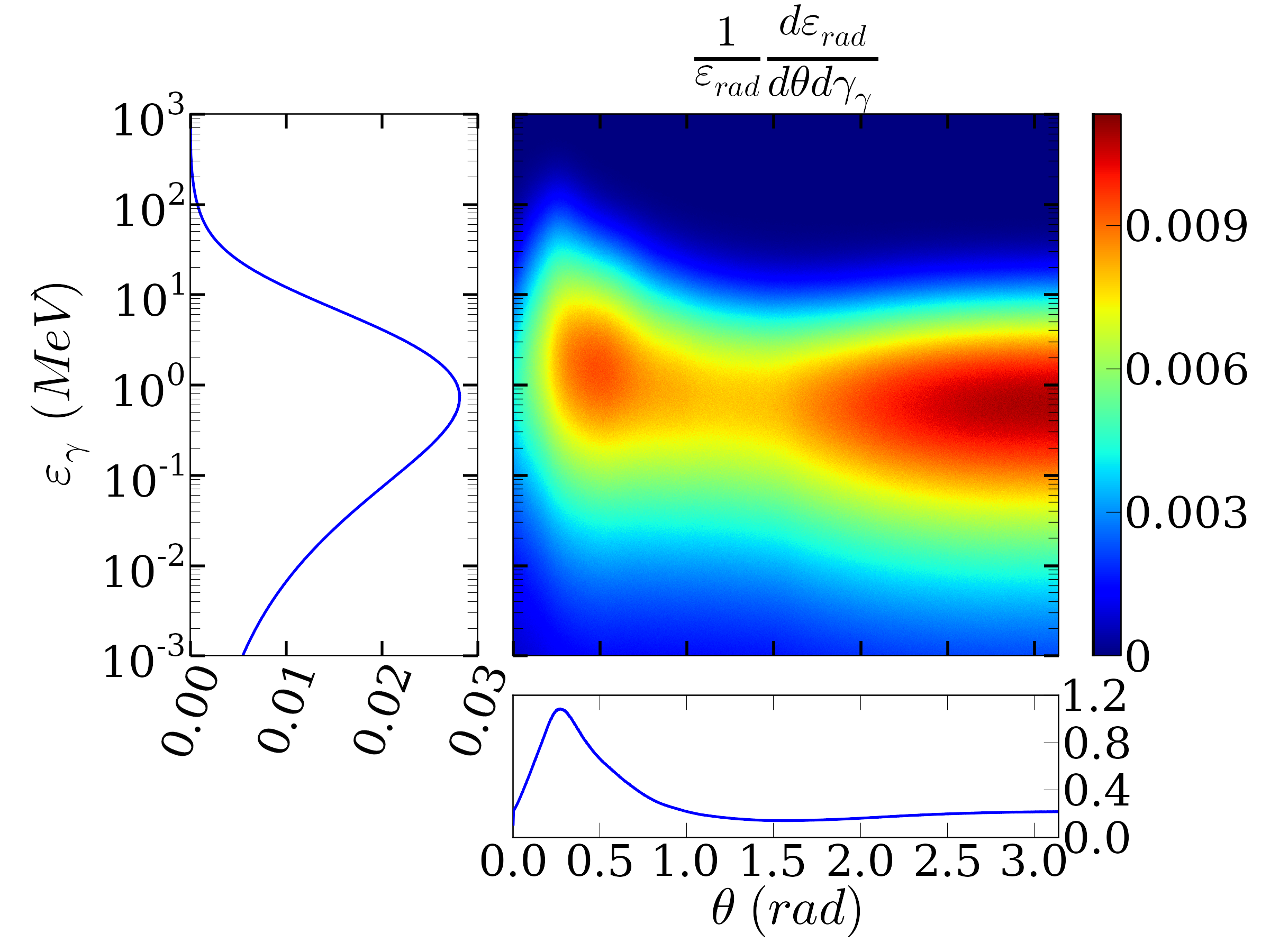}
\caption{\label{fig5} Spectral and angular distribution of the emitted photon energy in the interaction of a laser pulse at $10^{23}$ W/cm$^2$, in the case of gas target, $n_e=4 n_c$, $80 \mu$m.}
\end{figure}

At the end of simulation more than 90\% of the laser energy has been absorbed and 45\% has been transferred to high--energy photons. The time--integrated photon spectrum in energy $\varepsilon_{\gamma}$ and emission angle $\theta$ is shown in Fig. \ref{fig5}. The maximum emission is located around $28^\circ$ and the angular width corresponds to $15^\circ$. The forward emitted energy photon is 47 J (70 \% of the radiated energy) in a range 0.1--10 MeV. A brilliance is equal to $10^{16}$ photons/sr/$\mathrm{mm}^2$/s/0.1 bandwidth and a brightness is 0.4 J/MeV/sr. The number of forward emitted photons of an energy up to 1 MeV is equal to $2\times 10^{13}$. The average photon energy is equal to 2.7 MeV. 

By colliding two such photon beams at the source according to Eq.(\ref{eq5}) one may create $10^8$--$10^9$ pairs. At a distance $R=500\ \mu$m from the source, the beam average radius is $ R_{{\gamma}R} \sim 200\ \mu$m and the photon density $6 \times 10^{19}$ cm$^{-3}$. The expected yield of BW pair according to Eq.(\ref{eq6}) is $10^3$--$10^4$ similar to the case of an aluminium target. The focal spot is larger in case of a gas target but the pulse duration is longer.

An estimate of the background pair yield due to the Bethe--Heitler mechanism, is of the order of $10^8$ ($\sim 10^7$ forward emitted). This is similar for the aluminium target. A suppression of the pair production due to a smaller ion charge is compensated by the difference in the photon spectrum, the total radiated energy, and the fact that all photons (forward and backward emitted) contribute to the BH process. Therefore, for the detection of the BW process one needs to separate the photon--photon interaction zone from the source targets. The collision of photon beams at an angle $\theta \sim 90^\circ$ offers another important advantage for the BW pairs detection, as both, electron and positron, will be emitted in the preferential bisection direction. This may allow a better signal--to--noise ratio even for a large total number of background pairs. 

\section{Conclusion}\label{sec5}
Qualitative estimates and numerical simulations show that about $10^8$ BW pairs can be produced with existing sources of MeV photons. This number is comparable with the expected number of BH pairs. A separation of the BW and BH processes can be achieved by placing the photon collision zone far from the source. About $10^4$ pairs can be generated at the distance R=500 $\mu$m from the source. This number is comparable to the pair number given in \cite{Pike_2014}, however, with a much better control of background processes. The choice of the distance and a crossing angle give a possibility to optimize the signal--noise ratio. Two suitable $\gamma$--ray sources, the Bremsstrahlung and synchrotron sources are offering a good conversion efficiency. Although the expected number of BW pairs is sufficiently high, the major challenge is to discriminate them from other pairs created by the Trident and Bethe--Heitler processes in the photon source target. A shield could be designed to eliminate high energy photons and 
pairs production processes \cite{Sarri_Nat_2015,Sarri_PRL_2013}. Moreover, to deflect the background pairs, a strong magnetic field (100 Tesla) could be used \cite{Joao_2015}. A spatial separation of the photon--photon interaction zone is a promising way for the detection of the BW pairs emitted in the preferential direction. 

\begin{acknowledgments}
We acknowledge the financial support from the French National Research Agency (ANR) in the frame of "The Investments for the Future" Programme IdEx Bordeaux -- LAPHIA (ANR-10-IDEX-03-02) -- Project TULIMA. This work is partly supported by the Aquitaine Regional Council (project ARIEL). 
\end{acknowledgments}

\end{document}